\documentclass[twocolumn,superscriptaddress,showpacs,aps,prb]{revtex4}

\usepackage{amssymb}
\usepackage{amsmath}
\usepackage{amsfonts}
\usepackage{mathrsfs}

\usepackage{bm}
\usepackage{graphicx}

\newcommand{\ti}{\Tilde}
\newcommand{\nl}{\nonumber \\}

\newcommand{\Sec}[1]{Sec.\;\ref{#1}}

\newcommand{\be}{\begin{equation}}
\newcommand{\ee}{\end{equation}}
\newcommand{\bea}{\begin{eqnarray}}
\newcommand{\eea}{\end{eqnarray}}
\newcommand{\bal}{\begin{align}}
\newcommand{\eal}{\end{align}}
\newcommand{\bsube}{\begin{subequations}}
\newcommand{\esube}{\end{subequations}}
\newcommand{\Eq}[1]{Eq.\,(\ref{#1})}
\newcommand{\Eqs}[1]{Eqs.\,(\ref{#1})}
\newcommand{\Fig}[1]{Fig.\,\ref{#1}}
\newcommand{\dg}{\dagger}
\newcommand{\la}{\langle}
\newcommand{\ra}{\rangle}

\begin{document}

\title{Noise spectrum of quantum transport through quantum dots: \\
 a combined effect of non-Markovian and cotunneling processes}

\author{Jinshuang Jin } \email{hznu.jin@gmail.com}

\affiliation{ Department of Physics, Hangzhou Normal University,
  Hangzhou 310036, China}

\author{Wei-Min Zhang} \email{wzhang@mail.ncku.edu.tw}
\affiliation{Department of Physics and Center for Quantum Information Science,
National Cheng Kung University, Tainan 70101, Taiwan}

\author{Xin-Qi Li}
\affiliation{ Department of Physics, Beijing Normal University,
  Beijing 100875, China}

\author{YiJing Yan}
\affiliation{Department of Chemistry, Hong Kong University
   of Science and Technology, Kowloon, Hong Kong}

\date{\today}

\begin{abstract}

Based on our recently developed quantum transport theory in term of
an exact master equation, the corresponding particle-number resolved
($n$-resolved) master equation and the related shot noise spectrum
formalism covering the full frequency range are constructed.
We demonstrate that the noise spectra of transport current through
single quantum dot and double quantum dots show characteristic steps
and/or peak-dips in different tunneling frequency regimes through
tuning the applied bias voltage and/or gate voltage at low
temperatures.
The peak-dips crossing the tunneling resonant frequencies is a
combination effect of non-Markovian and cotunneling processes. These
voltage-dependent tunneling resonance characteristics can be
utilized to effectively modulate the internal Rabi resonance
signature in the noise spectrum.

\end{abstract}

\maketitle

\section{Introduction}
\label{thintro}

 Shot noise of non-equilibrium current fluctuations
contains rich information beyond the average
current.\cite{Bla001,Naz03} It  has
stimulated great interest in recent years both in theory and
experiment.\cite{Sun9910748,Eng04136602,Ent07193308,Fli08150601,Luo07085325,Don08033532,%
Saf03136801,Kie07206602,Zha07036603,Rot09075307,Jin11053704}
The evaluation of shot noise depends on the development of
non-equilibrium quantum transport theory. Conventional approaches to
quantum transport include the Landauer-B\"{u}ttiker scattering
matrix theory,\cite{Bla001,But91199} the non-equilibrium Green's
function formalism,\cite{Che914534,Hun9314687} and the real-time
diagrammatic technique.\cite{Bra06075328,Sch9418436} Many
interesting problems such as super- and sub-Poissonian noises and
their origins  have been
addressed.\cite{Bla001,Thi05045341,Dav929620,Fli08150601} However,
most of the investigations focused only on the zero- or
low-frequency regions.
The evaluation of noise spectrum over the full frequency range is
largely lacking and remains challenging with these approaches.
 Another commonly used method in quantum transport is the
quantum master equation approach. In a certain sense it is simpler
and more straightforward than the approaches mentioned above.
\cite{Luo07085325} It has the advantage of generality for different
scattering processes being handled in a unified manner, even the
transient dynamics under the time-dependent bias voltage.
\cite{Li07075114,Jin08234703,Jin10083013}

Quantum transport theory based on master equation has been developed
rapidly and studied extensively in recent years.
\cite{Sun9910748,Eng04136602,Fli08150601,Luo07085325,Don08033532,Jin11053704,
Kie06033312,Li05205304,Ent07193308,Bra06026805}
However, most of work were based on a certain perturbative master
equation description, which is valid mainly in Markovian dynamics
dominated regime with sequential tunneling processes involving only
one electron tunneling events at a time.
\cite{Sun9910748,Eng04136602,Fli08150601,Luo07085325,Don08033532,Jin11053704,
Kie06033312,Li05205304,Ent07193308}
Very recently, on the basis of Feynman-Vernon influence functional
technique,\cite{Fey63118} one of us has obtained an exact master
equation for nanostructures,\cite{Tu08235311} from which we have
also established an exact non-equilibrium quantum transport theory
 for noninteracting electronic systems.\cite{Jin10083013}
This new non-equilibrium quantum transport theory is applicable for
arbitrary voltage, arbitrary temperature and arbitrary
system-reservoir coupling strength.
Quantum transport based on this exact treatment can deal with all
the non-Markovian tunneling processes accompanied with not only
sequential tunneling but also cotunneling involving two or more
electron tunneling events at the same time. \cite{Fra01878,Ave92}

In this paper, we will construct the corresponding master equation
for the reduced density matrix conditioned on the electron number
passed through the tunnel junction, i.e. the particle-number
resolved ($n$-resolved) master equation. Together with the
MacDonald's formula, we give the description for non-Markovian noise
spectrum over the full frequency range.
All relevant formulations will be summarized and developed in
\Sec{thform} together with the detailed numerical method presented
in Appendix. This formalism is certainly applicable to arbitrary
bias voltage, temperature and system-reservoir coupling strength for
all the non-Markovian processes.
In \Sec{thappl}, based on the present formulation, we study the
quantum transport setup for the shot noise of current through two
model systems. The first model is a single one-level quantum dot
system, operated in the conventional sequential tunneling regime
when only the average current is concerned. From the noise spectrum,
however, we find that there emerge peak-dips in the noise spectrum
as the temperature decreases. Another example is a coupled double
quantum dots system, with one level in each dot. The resulting noise
spectrum shows both the coherent Rabi resonance of the dots system
and the external field-dependent tunneling characteristic resonant
structure. We observe that the coherent Rabi resonance is
insensitive to the temperature, while the external field-dependent
tunneling resonant structure has the very interesting
temperature-dependent phenomena. In particular, the external
field-dependent tunneling resonance characteristics is manifested
with either a peak-dip or a step at low temperature, depending on
the corresponding tunneling regime defined by the applied bias
and/or gate voltage.
Consequently, by varying the applied voltage
one can tune the tunneling resonance on
top of the Rabi resonance, thus effectively
modulate the coherent signature in noise spectrum.
Finally, a summary is presented in \Sec{thsum}.

\section{$n$-resolved master equation and noise spectrum}
\label{thform}

\subsection{Exact master equation and transient transport current}
\label{thformA}

 Let us start with a brief outline
of the recently developed exact master equation and the
non-equilibrium quantum transport
theory.\cite{Tu08235311,Jin10083013} The quantum transport studied
with a nano-device consists of such as quantum dots interacting with
the electron reservoirs (source and drain). The total Hamiltonian of
the system has three parts: $H_T=H_S+H_B+H'$. The central system of
quantum dots part assumes $H_S=\sum_\mu \epsilon_{\mu\nu} a^\dg_\mu
a_\nu$, where $a^\dg_{\mu}$ ($a_{\mu}$) is the creation
(annihilation) operator of electron at the specified state of energy
$\epsilon_{\mu\mu}\equiv\epsilon_\mu$, while the off-diagonal
$\epsilon_{\mu\nu}$ is the hopping integral. The Hamiltonian of
free-electron reservoirs reads $H_B=\sum_{\alpha k}\epsilon_{\alpha
k}c^\dg_{\alpha k}c_{\alpha k}$, with $c^\dg_{\alpha k}$ being the
creation operator for the electron in the reservoir $\alpha=L,R$.
The system-reservoirs coupling  for electron tunneling is given by
\be\label{Hprime}
 H' = \sum_{\alpha k \mu}V_{\alpha k\mu}a^\dg_\mu  c_{\alpha k}
    + {\rm H.c.}
\ee where $V_{\alpha k\mu}$ is the tunneling coefficient or the
tunneling amplitude between the system and the reservoir $\alpha$.
The electron-electron interaction has been ignored here.
Then the reduced density matrix of the central system
$\rho(t)\equiv{\rm tr}_B[\rho_{\rm T}(t)]$, i.e., the trace of total
density matrix over reservoirs degrees of freedom, is governed by
the exact master equation\cite{Tu08235311,Jin10083013}
\begin{align}
 \dot\rho(t)&=-i[{\widetilde H}_S(t),\rho(t)] -\sum_{\alpha\mu \nu}
 \widetilde{\bm \gamma}_{\alpha\mu \nu}(t)
  \big[a^{\dag}_{\mu},a_{\nu}\rho(t)+\rho(t) a_{\nu} \big]
\nl&\quad
  - \sum_{\alpha\mu \nu}\!{\bm\gamma}_{\alpha\mu \nu}(t)
  \Big(\frac{1}{2} a^{\dag}_{\mu}a_{\nu}\rho(t)
     \!+\! \frac{1}{2}\rho(t) a^{\dag}_{\mu}a_{\nu}
     \!-\! a_{\nu}\rho(t) a^{\dag}_{\mu}
  \Big).
\label{QME_LB}
\end{align}
where the first term represents the Liouville equation (the unitary
part) of the system with the renormalized Hamiltonian ${\widetilde
H}_S(t)=\sum_{\mu\nu}\ti\epsilon_{\mu\nu}a^\dg_\mu a_\nu$ and the
renormalized energy levels $ \ti{\bm
\epsilon}(t)=\bm\epsilon(t)-\frac{i}{2}\sum_\alpha\big[\bm\kappa_\alpha(t)-
\bm\kappa^\dg_\alpha(t)\big]$. The other two terms (the non-unitary
part) in \Eq{QME_LB} describe the dissipation and decoherence
dynamics of the system with the time-dependent dissipation and
fluctuation coefficients $\bm\gamma_\alpha(t)\equiv
\frac{1}{2}[\bm\kappa_\alpha(t)+\bm\kappa^\dg_\alpha(t)]$ and
$\widetilde{\bm\gamma}_\alpha(t)\equiv\bm\lambda_\alpha(t)+\bm\lambda^\dg_\alpha(t)$,
where
 $\bm \kappa_\alpha(t)$ and $\bm \lambda_\alpha(t)$ are given by,
\bsube \label{uu-sol}
\begin{align}
\bm \kappa_\alpha(t)= & \int_{t_0}^{t}d\tau \bm
g_\alpha(t,\tau)\bm u(\tau) [\bm u(t)]^{-1} ,  \label{uu-2} \\
\bm \lambda_\alpha(t) = & \int_{t_0}^{t} d\tau \big\{ \bm
g_{\alpha}(t,\tau) \bm v(\tau) -\bm
g_{\alpha}(t,\tau)\bar{\bm u}(\tau) \big\}
 -\bm\kappa_\alpha(t) \bm v(t) . \label{uu-3}
\end{align}
\esube and  $\bm u(\tau)$ and $\bm v(\tau)$ are indeed directly
related to the Keldysh's non-equilibrium Green functions and they
obey the following dissipation-fluctuation integrodifferential
(Dyson) equations: \cite{Tu08235311,Jin10083013}
 \bsube\label{uv-eq}
\begin{align}
&\dot{\bm u}(\tau)+i\bm \epsilon(\tau) {\bm u}(\tau) +  \sum_\alpha
\int_{t_0}^{\tau } d\tau' \bm g_\alpha (\tau,\tau') {\bm
u}(\tau')=0,
\label{ut-eq}
\\
&\dot{\bm  v}(\tau)+i\bm \epsilon (\tau)\bm v (\tau)  + \sum_\alpha
\int_{t_0}^{\tau } d\tau' \bm g_\alpha (\tau,\tau') \bm v(\tau')
\nl&\quad\quad\quad\quad\quad\quad
=\sum_\alpha \int_{t_0}^{t }d\tau'
      \bm{\widetilde g}_\alpha (\tau,\tau')\bar{\bm u}(\tau'),
\label{vt-eq}
\end{align}
\esube subject to the boundary conditions $\bm u(t_0)=\bm 1$, and
$\bm v(t_0)=0$. The integral kernels in the above equations, $\bm
g_\alpha(\tau,\tau')$ and $\bm{\widetilde g}_\alpha(\tau,\tau')$,
are defined by $\bm g_{\alpha \mu\nu}(\tau,\tau')
 =\sum_{k} V_{\alpha k\mu}V^*_{ \alpha k\nu}
   e^{-i\epsilon_{\alpha k}(\tau-\tau')}
$ and $\widetilde{\bm g}_{\alpha \mu\nu}(\tau,\tau')
 =\sum_{k} V_{\alpha k\mu}V^*_{ \alpha k\nu}
 f_\alpha(\epsilon_{\alpha k})
    e^{-i\epsilon_{\alpha k}(\tau-\tau')}$,
which depict all the non-Markovian memory processes of electrons
through the tunnelings between the dot system and the leads, and
$f_\alpha (\epsilon_{\alpha k})= 1/(e^{\beta_\alpha
(\epsilon_{\alpha k}-\mu_\alpha)} -1)$ is the Fermi distribution
function of the lead $\alpha$ in the initial equilibrium state.
Introducing the spectral density functions of the $\alpha$-lead
coupled with the dot system: $\bm \Gamma_{\alpha}(\omega)\equiv
\{\Gamma_{\alpha\mu\nu}(\omega)=2\pi \sum_k
 V_{\alpha  k\mu}V^{\ast}_{\alpha k\nu}
 \delta(\omega-\epsilon_{\alpha k})\}$, the memory kernels are reduced to
 \bsube \label{ggbeta}
 \begin{align}
 \bm g_{\alpha }(\tau,\tau')
 &= \int \frac{d\omega}{2\pi} \bm \Gamma_{\alpha}(\omega)
 e^{-i\omega(t-\tau)},
 \\
 \widetilde{\bm g}_{\alpha }(\tau,\tau')
 &= \int \frac{d\omega}{2\pi} f_\alpha(\omega)\bm\Gamma_{\alpha}(\omega)
 e^{-i\omega(t-\tau)}.
 \end{align}
 \esube
 Through out this work, we have set $e=\hbar=1$.

 The transient current of the $\alpha$-lead is obtained as\cite{Jin10083013}
\begin{align}
 I_\alpha(t) &= - 2{\rm Re}\!\int_{t_0}^{t}\!d\tau
  {\rm Tr}
\Big\{
    {\bm g}_{\alpha}(t-\tau)\bm v(\tau)
   -\widetilde{\bm g}_{\alpha}(t-\tau)\bar{\bm u}(\tau)
\nl&\qquad\qquad\qquad\ \
   +{\bm g}_\alpha(t,\tau)\bm u(\tau)\bm \varrho(t_0){\bm u}^\dg(t)
\Big\},
\label{curr}
\end{align}
with $\bar{\bm u}(\tau)={\bm u}^\dg(t-\tau)$ and ${\bm\varrho}(t)$
being the single particle reduced density matrix,
$\varrho_{\mu\nu}(t)\equiv {\rm tr}_s\big[a^\dg_\nu
a_\mu\rho(t)\big]$, satisfying $\bm \varrho(t)=\bm v(t)+ \bm
u(t)\bm\varrho(t_0)\bm u^\dag(t)$. The above current expression can
easily recover the Landuer-B\"{u}ttiker formula obtained by
scattering approach and the Keldysh's non-equilibrium Green function
formulation, as detailed in Ref.\,\onlinecite{Jin10083013}.


\subsection{ $n$-resolved quantum master equation}

Before constructing the $n$-resolved master equation, we should
point out that the master equation of \Eq{QME_LB} was derived
exactly for noninteracting quantum dot systems so that all the
electron tunneling processes in such systems, including sequential
tunneling and cotunneling processes together with all the
non-Markovian memory effects, are fully taken into account.
For noninteracting systems, the cotunneling processes refer to the
tunnelings involving two and more electrons through the junctions
simultaneously. In the conventional master equation in terms of
perturbation expansion, one may expect that the sequential tunneling
and cotunneling processes should be described separately by one pair
and two or more pairs of electron creation and annihilation
operators, respectively, in the non-unitary part of the master
equation. However, for noninteracting systems, the master equation
derived from the perturbation expansion up to the second order of
the tunneling coefficients, i.e. $V_{\alpha k\mu}$ in \Eq{Hprime},
has indeed the same operator structure as that of the exact master
equation of \Eq{QME_LB}.\cite{Tu08235311} The only difference
between the second-order ($2^{\rm nd}$-order) perturbative master
equation and the exact master equation is manifested in the
determination of the dissipation and fluctuation coefficients in the
corresponding master equations. In the exact master equation of
\Eq{QME_LB}, these dissipation and fluctuation coefficients, i.e.,
$\gamma_{\alpha\mu\nu}(t)$ and $\ti\gamma_{\alpha\mu\nu}(t)$, are
determined by the Green functions $\bm u(\tau)$ and $\bm v(\tau)$
which satisfy the exact dissipation-fluctuation integrodifferential
equations of \Eq{uv-eq}. The non-Markovian memory effect and the
cotunneling processes described in the exact master equation are
fully characterized by the non-local integral kernels in \Eq{uv-eq},
i.e. the Dyson equations through the iteration to all orders of the
tunneling coefficients. Truncating \Eq{uv-eq} to the $2^{\rm
nd}$-order perturbation of the tunneling coefficients, all these
time-dependent dissipation and fluctuation coefficients are reduced
to the coefficients in the $2^{\rm nd}$-order perturbative master
equation that can be derived directly from the quantum Lioulille
equation in the perturbation expansion approach. Since the $2^{\rm
nd}$-order perturbative master equation can only describe the
sequential tunneling and Markovian process, it indicates that the
cotunneling processes and non-Markovian effect in noninteracting
systems are fully determined by the time-dependent dissipation and
fluctuation coefficients from the solution of \Eq{uv-eq}, rather
than the operator structure of the master equation. The operator
structure of the $2^{\rm nd}$-order perturbative master equation is
exactly the same as that of the exact master equation for
noninteracting systems. The detailed derivation of this connection
between the exact master equation and the $2^{\rm nd}$-order
perturbative master equation has been given in
Ref.\,\onlinecite{Tu08235311}.

Based on such a connection between the exact master equation and the
$2^{\rm nd}$-order perturbative master equation we just discussed
above, we can construct the corresponding $n$-resolved master
equation from \Eq{QME_LB} for the conditioned reduced density
operator $\rho^{(n_\alpha)}$, from the $n$-resolved $2^{\rm
nd}$-order perturbative master equation that has been derived
explicitly from the $2^{\rm nd}$-order perturbative master
equation.\cite{Li05205304} $\rho^{(n_\alpha)}$ is conditioned on the
registered number $n_\alpha$ of electrons having passed through the
tunnel junction between the specified $\alpha$--lead and the central
dots system.
As it has been shown in Ref.\,\onlinecite{Li05205304}, the resulting
$n$-resolved $2^{\rm nd}$-order perturbative master equation only
modifies the corresponding non-unitary operator structure of the
$2^{\rm nd}$-order perturbative master equation, namely, it replaces
only the registered jump terms $a^\dg_\mu \rho a_\nu$ and $a_\mu
\rho a^\dg_\nu$, associating with the specified $\alpha$-lead in the
master equation by $a^\dg_\mu \rho^{(n_\alpha+1)} a_\nu$ and $a_\mu
\rho^{(n_\alpha-1)} a^\dg_\nu$, respectively. Other terms remain the
same but for $\rho^{(n_\alpha)}$.
Since the exact master equation for noninteracting systems has the
same operator structure of the $2^{\rm nd}$-order perturbative
master equation, by analogy, we obtain the corresponding
$n$-resolved master equation of the exact master equation of
\Eq{QME_LB}:
 \begin{widetext}
 \begin{align}\label{nME}
 \dot \rho^{(n_\alpha)}(t)\!
&= -i[{\widetilde H}_S,\rho^{(n_\alpha)}(t)] 
  -\sum_{\mu\nu}\!
   \widetilde{\bm\gamma}_{\mu\nu}(t)
   \big[
     a^\dg_{\mu} a_\nu \rho^{(n_\alpha)}(t)
    -\rho^{(n_\alpha)}(t) a_\nu a^\dg_{\mu}
  \big]-\sum_{\mu\nu}\!{\bm\gamma}_{\mu\nu}(t)
   \big\{a^\dg_{\mu} a_\nu, \rho^{(n_\alpha)}(t)
  \big\}
\nl&\quad
 +\sum_{\mu\nu}\!
  \big[\widetilde{\bm\gamma}_{\alpha'\!\mu\nu}(t)
    \!+\!2\bm\gamma_{\alpha'\!\mu\nu}(t)\big]
       a_\nu \rho^{(n_\alpha)}(t) a^\dg_{\mu}
   -\widetilde{\bm\gamma}_{\alpha'\!\mu\nu}(t)
     a^\dg_\mu\rho^{(n_\alpha)}(t) a_\nu
  \big]
\nl&\quad
 +\sum_{\mu\nu}
   \big[\widetilde{\bm\gamma}_{\alpha\mu\nu}(t)+2\bm\gamma_{\alpha\mu\nu}(t)\big]
     a_\nu \rho^{(n_\alpha-1)}(t) a^\dg_{\mu}
  -\sum_{\mu\nu} \widetilde{\bm\gamma}_{\alpha\mu\nu}(t)
     a^\dg_\mu\rho^{(n_\alpha+1)}(t) a_\nu.
 \end{align}
  \end{widetext}
for keeping track the number of electrons tunneling only through one
specific lead, e.g., the $\alpha$-lead with $\alpha'\neq\alpha$, and
 \begin{widetext}
 \begin{align}\label{nME2}
  \dot \rho^{(n_L, n_R)}(t)\!
&= -i[{\widetilde H}_S,\rho^{(n_L, n_R)}(t)] 
  -\sum_{\mu\nu}\!
   \widetilde{\bm\gamma}_{\mu\nu}
   \big[
     a^\dg_{\mu} a_\nu \rho^{(n_L, n_R)}(t)
    -\rho^{(n_L, n_R)}(t) a_\nu a^\dg_{\mu}
  \big]
  \nl&\quad
  -\sum_{\mu\nu}\!{\bm\gamma}_{\mu\nu}
   \big\{a^\dg_{\mu} a_\nu, \rho^{(n_L, n_R)}(t)
  \big\}
\nl&\quad
 +\sum_{\mu\nu}\!
  \big[(\widetilde{\bm\gamma}_{L\mu\nu}
    \!+\!2\bm\gamma_{L\mu\nu})
       a_\nu \rho^{(n_L-1, n_R)}(t) a^\dg_{\mu}
   -\widetilde{\bm\gamma}_{L\mu\nu}
     a^\dg_\mu\rho^{(n_L+1, n_R)}(t) a_\nu
  \big]
\nl&\quad
 +\sum_{\mu\nu}
   \big(\widetilde{\bm\gamma}_{R\mu\nu}+2\bm\gamma_{R\mu\nu}\big)
     a_\nu \rho^{(n_L,n_R-1)}(t) a^\dg_{\mu}
  -\sum_{\mu\nu} \widetilde{\bm\gamma}_{R\mu\nu}
     a^\dg_\mu\rho^{(n_L,n_R+1)}(t) a_\nu,
 \end{align}
  \end{widetext}
for conditionally keeping track the number of electrons tunneling
through both the right and the left leads. Here we omit the time index in
the time-dependent coefficients and have also defined
$\widetilde{\bm\gamma}_{\mu\nu}\equiv
\sum_{\alpha=L,R}\widetilde{\bm\gamma}_{\alpha\mu\nu}$.
Both the $n$-resolved master equations of \Eq{nME} and \Eq{nME2}
will be used to derive the noise spectrum of an individual lead and
the cross-correlation noise spectrum.

The same as the connection between the exact master equation and the
$2^{\rm nd}$-order perturbative master equation discussed in the
beginning of this subsection, the $n$-resolved master equation,
\Eq{nME} and \Eq{nME2}, also has the same operator structure as that
of the $n$-resolved $2^{\rm nd}$-order perturbative master
equation.\cite{Li05205304} The only difference is the dissipation
and fluctuation coefficients in these master equations. The
dissipation and fluctuation coefficients in the $n$-resolved $2^{\rm
nd}$-order perturbative master equation involve only sequential
tunneling process and the Markovian dynamics.\cite{Li05205304}
However, the time-dependent dissipation and fluctuation coefficients
in \Eq{nME} and \Eq{nME2} that are determined by the exact
non-equilibirum Green functions of \Eq{uv-eq} for noninteracting
systems have taken into account simultaneously the sequential
tunneling and cotunneling processes, so does for the non-Markovian
dynamics.
 One may ask why the states such as $\rho^{(n_\alpha\pm2)}$,
$\rho^{(n_L-1,n_R+1)}$ and $\rho^{(n_L+1,n_R-1)}$, etc. which are
expected to be directly related to the cotunneling processes for the
electron registered detector in the perturtive expansion do not
occur in \Eq{nME} and \Eq{nME2}. This must be the same reason as in
the case of the master equation itself. In terms of the perturbatve
expansion, the master equation up to the higher order contains the
dissipation and fluctuation terms involving two and more pairs of
electron operators sandwiching the reduced density matrix that
describe the cotunneling processes. But the exact master equation
does not contain such terms. This is because the cotunneling
processes has been all switched into the dissipation and fluctuation
coefficients determined by the exact Dyson equations. Although we
have not provided a rigorous derivation of \Eq{nME} and \Eq{nME2},
it should be the same reason that the $n$-resolved master equation,
\Eq{nME} and \Eq{nME2}, can maintain the simple operator form
because the time-dependent dissipation and fluctuation coefficients
determined by the exact nonequilibrium Green functions can address
all the sequential tunneling and cotunneling processes as the
non-Markovian memory dynamics.

As a justification and also a self-consistent check, from the
$n$-resolved master equation of \Eq{nME} with the identity
$\rho(t)=\sum_{n_\alpha} \rho^{(n_\alpha)}(t)$, we can easily
reproduce the exact master equation of \Eq{QME_LB}. The exact
current expression of \Eq{curr} can also be reproduced from the
$n$-resolved master equation of \Eq{nME} as follows. With the
knowledge of $\rho^{(n_\alpha)}(t)$, the tunneling electrons
distribution can be readily evaluated via $P(n_\alpha,t)={\rm
Tr}_S\rho^{(n_\alpha)}(t)$. This is the key quantity for full
counting statistics.\cite{Bag03085316,Kie06033312,Gus06076605} The
$m^{\rm th}$-moment of transport is just $\la
n^m_\alpha(t)\ra\equiv\sum_{n_\alpha} n^m_\alpha P(n_\alpha,t)$,
from which all transport properties can be obtained. For instance,
the measured current which is related to the rate of the first
moment, is given by
$ I_\alpha(t)=-\frac{d}{dt}\la n_\alpha(t)\ra$. Using the
$n$-resolved master equation of \Eq{nME}, it is also easy to
reproduce the exact expression of the transient transport current of
\Eq{curr}. The capability of recovering the exact master equation of
\Eq{QME_LB} and the exact transport current formula of \Eq{curr}
from \Eq{nME} ensures that the $n$-resolved master equations we
constructed here is the current one for noninteracting systems.
Thus the current-current correlation noise spectrum which is related
to the $2^{\rm nd}$-order cumulant can be directly evaluated now
from \Eq{nME} and \Eq{nME2}, which will be presented in the
following.

\subsection{Noise spectrum expression}
\label{thnoise}
 We consider now the current noise spectrum,
$S(\omega)= \mathscr{F}\{\la \delta I(t)\delta I(0)\ra_s\}$,
i.e.,
the full Fourier transformation (denoted by $\mathscr{F}$) of the
fluctuating current-current correlation function that is symmetrized.
For the total circuit current $I(t)=aI_L(t)-bI_R(t)$,
which is typically the measured quantity in most experiments,\cite{Bla001} with the
coefficients satisfying $a+b=1$ related to the
symmetry of the transport setup (e.g., junction capacitances),
the circuit noise spectrum is
$S(\omega)=a^2 S_L(\omega)+b^2 S_R(\omega)-2ab S_{LR}(\omega)$.
\cite{Bla001,Luo07085325,Eng04136602}

The noise spectrum at individual lead $\alpha=L$ or $R$
is given by the MacDonald's formula,
$S_{\alpha}(\omega)\equiv  S_{\alpha\alpha}(\omega)
 =2\omega\int_0^{\infty}dt\, \sin(\omega t)
\frac{d}{dt}\big[\langle n_{\alpha}^2(t)\rangle-(\bar I_\alpha t)^2\big]$.
With the help of \Eq{nME}, the involved quantity $\langle n_{\alpha}^2(t)\rangle$
satisfies
\begin{align}\label{nkalpha}
\frac{d}{dt}\la n^2_\alpha(t)\ra
&=
 2\,{\rm tr}\left[\big(\widetilde{\bm\gamma}_{\alpha}+\bm\gamma_{\alpha}\big)
  \bm{\mathcal {N}}_\alpha
 +\widetilde{\bm\gamma}_{\alpha} \bar{\bm{\mathcal {N}}}_\alpha\right]
\nl&\quad
 +{\rm tr}\left[\big(\widetilde{\bm\gamma}_{\alpha}+\bm\gamma_{\alpha}\big)
  \bm\varrho
 -\widetilde{\bm\gamma}_{\alpha} \bar{\bm\varrho}\right],
\end{align}
where $\bar{\bm\varrho}=1-{\bm\varrho}$ is the single hole particle
reduced density matrix, while
 $\mathcal {N}_{\alpha\mu\nu}\equiv{\rm tr}_s [a^\dg_\nu a_\mu\hat N^\alpha ]$ and
 $\bar{\mathcal {N}}_{\alpha\mu\nu}\equiv{\rm tr}_s [ a_\mu a^\dg_\nu\hat N^\alpha]$
are related to the number operator
$\hat N^\alpha(t)\equiv\sum_{n_\alpha}n_\alpha\rho^{(n_\alpha)}(t)$.
It satisfies
\begin{align}\label{hatNt}
 \frac{d\hat N^\alpha}{dt}
&={\cal M}\hat N^\alpha
 +{\rm tr}\left[\big(\widetilde{\bm\gamma}_{\alpha}
                 +\bm\gamma_{\alpha}
                \big){\bm \varrho}
          \right]
 +{\rm tr}\left[\widetilde{\bm\gamma}_{\alpha} {\bm{\bar \varrho}}\right],
\end{align}
with ${\cal M}$ being the generator of Eq.~(\ref{QME_LB}) that is
recast as $\dot\rho={\cal M}\rho$.
 The initial conditions to \Eqs{nkalpha} and (\ref{hatNt}) are
 the stationary states without the electron tunneling,
 i.e., $\hat N_\alpha(0)=0$ and $\rho(0)=\rho^{\rm st}$.
We have
 \begin{align}\label{Salpha}
 S_\alpha(\omega)
&= 4\omega{\rm Im}\,\mathscr{L}\left\{
  {\rm tr}\big[\big(\widetilde{\bm\gamma}_{\alpha}
    +\bm\gamma_{\alpha}\big)\bm{\mathcal {N}}_\alpha
    +\widetilde{\bm\gamma}_{\alpha} \bar{\bm{\mathcal {N}}}_\alpha
 \big]\right\}
\nl&\quad
+2\,{\rm Re}\,\mathscr{L}\Big\{{\rm tr}
  \big[\big(\widetilde{\bm\gamma}_{\alpha}
    +\bm\gamma_{\alpha}\big)\bm\varrho^{\rm st}
    -\widetilde{\bm\gamma}_{\alpha}{ \bar{\bm\varrho}^{\rm st}}
  \big]\Big\},
\end{align}
where
$\mathscr{L}\{f(t)\}=\int^\infty_0 dt\,e^{i\omega t}f(t)$
denotes the Laplace transformation.

    The cross correlation noise spectrum is
given by\cite{Wan04153301,Don05066601,Jin11053704}
$S_{LR}(\omega)= {1\over 2}\mathscr{F}
\big\{\la \delta I_{L}(t)\delta I_{R}(0)\ra_s
+\la \delta I_{R}(t)\delta I_{L}(0)\ra_s\big\}
=2\omega\!\int^\infty_0\!\!{\mathrm d}t \sin(\omega t)
\frac{{\mathrm d}}{{\mathrm d}t}\!
\big[\la N_{L}(t)N_{R}(t)\ra
- (\bar I t)^2\big]$,
where $\la N_{L}(t)N_{R}(t)\ra={\rm tr}
\sum_{n_{L}n_{R}}n_{L}n_{R} \rho^{(n_{L}\!,n_{R})}(t)$.
Again, with the help of \Eq{nME2}, we obtain
\begin{align}\label{SLR}
S_{LR}(\omega)
&=
2\omega{\rm Im}\mathscr{L}\Big\{
{\rm tr}\big[\big(\widetilde{\bm\gamma}_{L}+\bm\gamma_{L}\big)
  \bm{\mathcal {N}}_R
 +\widetilde{\bm\gamma}
 _{L} \bar{\bm{\mathcal {N}}}_R\big]
 \nl&\quad\quad\quad\quad
  +{\rm tr}\big[\big(\widetilde{\bm\gamma}_{R}+\bm\gamma_{R}\big)
  \bm{\mathcal {N}}_L
 +\widetilde{\bm\gamma}_{R} \bar{\bm{\mathcal {N}}}_L\big]
 \Big\}.
 \end{align}

We thus have completed the expressions of the current noise spectrum, i.e.,
\Eq{Salpha} and \Eq{SLR}, which are the main results of the present work.
They can be further written in the compacted formalism as
\begin{align}\label{Sw-exact}
 S_{\alpha\alpha'}(\omega)
&= 2\omega{\rm Im}\mathscr{L}\Big\{
  {\rm tr}\big[\big(\widetilde{\bm\gamma}_{\alpha}+\bm\gamma_{\alpha}\big)
  \bm{\mathcal {N}}_{\alpha'}
 +\widetilde{\bm\gamma}_{\alpha} \bar{\bm{\mathcal {N}}}_{\alpha'}\big]
 \nl&\qquad\qquad\,
 +{\rm tr}\big[\big(\widetilde{\bm\gamma}_{{\alpha'}}+\bm\gamma_{{\alpha'}}\big)
  \bm{\mathcal {N}}_{\alpha}
 +\widetilde{\bm\gamma}_{{\alpha'}} \bar{\bm{\mathcal {N}}}_{\alpha}\big]
 \Big\}
\nl&\quad
  +2\delta_{\alpha\alpha'}{\rm Re}\mathscr{L}\Big\{
   {\rm tr}\big[\big(\widetilde{\bm\gamma}_{\alpha}+\bm\gamma_{\alpha}\big)
  \bm\varrho^{\rm st}
 -\widetilde{\bm\gamma}_{\alpha}
  \bar{\bm\varrho}^{\rm st}\big]\Big\}.
 \end{align}
It describes both the sequential tunneling and cotunneling processes
together with the non-Markovian memory effect involved in the
time-dependent dissipation and fluctuation coefficients
$\bm\gamma_\alpha(t)$ and $\widetilde{\bm\gamma}_\alpha(t)$ that
also determine the master equation of \Eq{QME_LB} and the
$n$-resolved master equation of \Eq{nME}. We will evaluate
explicitly the noise spectrum using \Eq{Sw-exact} in the coming
section with two widely adopted quantum transport model systems.
Since the time-dependence dissipation and fluctuation coefficients
are determined by the functions $\bm u(\tau)$ and $\bm v(\tau)$
which are governed by \Eq{uv-eq}, the key to understand the
cotunneling processes and the non-Markovian effect in the transient
transport current and the noise spectrum is to solve exactly the
integrodifferential equations of Eq.~(\ref{uv-eq}). The detailed
numerical method is presented in Appendix based on a
parameterization scheme.

As it is shown in Ref.\,\onlinecite{Tu08235311}, $\bm u(\tau)$ and
$\bm v(\tau)$ accounts for all orders in perturbative expansion. If
one solves Eq.~(\ref{uv-eq}) up to the second-order of the
system-reservoir couplings, the resulting master equation recovers
the $2^{\rm nd}$-order master equation in the perturbation theory
that has been studied widely in the literature, see for examples
Refs.\,\onlinecite{Luo07085325,Fli08150601,Jin11053704}.
The usual Markovian approximation is obtained under both the wide
band limit $\bm\Gamma_\alpha(\omega)=\bm\Gamma_\alpha$ with high
temperature, which leads to $\bm
g_\alpha(t-\tau)\rightarrow\frac{\Gamma_\alpha}{2}\delta(t-\tau)$
and $\bm {\widetilde
g}_\alpha(t-\tau)\rightarrow\frac{\Gamma_\alpha}{2}f_\alpha\delta(t-\tau)$.
Then all the time-dependent coefficients in the master equation
become constants, e.g., $\bm\gamma_\alpha(t)=\Gamma_\alpha$ and
$\widetilde{\bm\gamma}_\alpha(t)= -f_\alpha\Gamma_\alpha$, which
recovers the $n$-resolved Markovian master equation in
Ref.\,\onlinecite{Gur9615932} and also reproduces the solution given
in Ref.\,\onlinecite{Bag03085316} for resonant-level model.

Usually one believes that in the wideband limit and the high bias
voltage regime, the non-Markovian effect vanishes and the Markovian
master equation works. However, the non-Markovian memory effect is
very complicated, it relates to several factors together, such as
the temperature, the bias voltage, the spectral bandwidth, and the
system-reservoir coupling strength as details in
Ref.\,\onlinecite{Jin10083013} and in Appendix.
In Ref.\,\onlinecite{Jin10083013}, we showed that in the WBL, the
large bias voltage indeed weakens the non-Markovian effect, but does
not lead to a complete Markovian limit. At this regime, the
non-Markovian effect, which arise completely from the low
temperature and/or strong system-reservoir coupling strength, may
not be clearly manifested in the transport current itself and the
reduced density matrix, \cite{Jin10083013} but it becomes
significant in the noise spectrum of the current-current
fluctuation, \cite{Jin11053704,Eng04136602} as shown in the next
section.

\section{Numerical demonstrations}
\label{thappl}

\subsection{Single quantum dot system}
For simplicity, we consider in this section the situation of an
energy-independent spectral density of the leads, namely, the flat
band of $\Gamma_\alpha(\omega)=\Gamma_\alpha$ and denote
$\Gamma=\Gamma_L+\Gamma_R$. We set the symmetric junction
capacitances as $a=b=0.5$ and also the symmetric bias of
$\mu_L=-\mu_R=eV/2$ in the following studies.
 Let us start with a single one-level dot model system, with
$H_S=\varepsilon a^\dg a$. The average current of this system can be
evaluated as $\bar I=\Gamma_L\Gamma_R\int
\frac{d\omega}{2\pi}\frac{f_L(\omega)-f_R(\omega)}
{(\varepsilon-\omega)^2+(\Gamma/2)^2}$. Despite of its simplicity,
such a single quantum dot system has many interesting physical
phenomena and has been widely used as a single electron transistor.
Here we operate the system in the large voltage regime,
 $\mu_L>\varepsilon>\mu_R$,
which is also considered as the conventional sequential tunneling
regime if the average current is concerned. However, we will see
below that in this case a striking non-Markovian character,
manifested by a peak-dip, will emerge in the current noise spectrum.

\begin{figure}
\includegraphics*[width=0.9\columnwidth,angle=0]{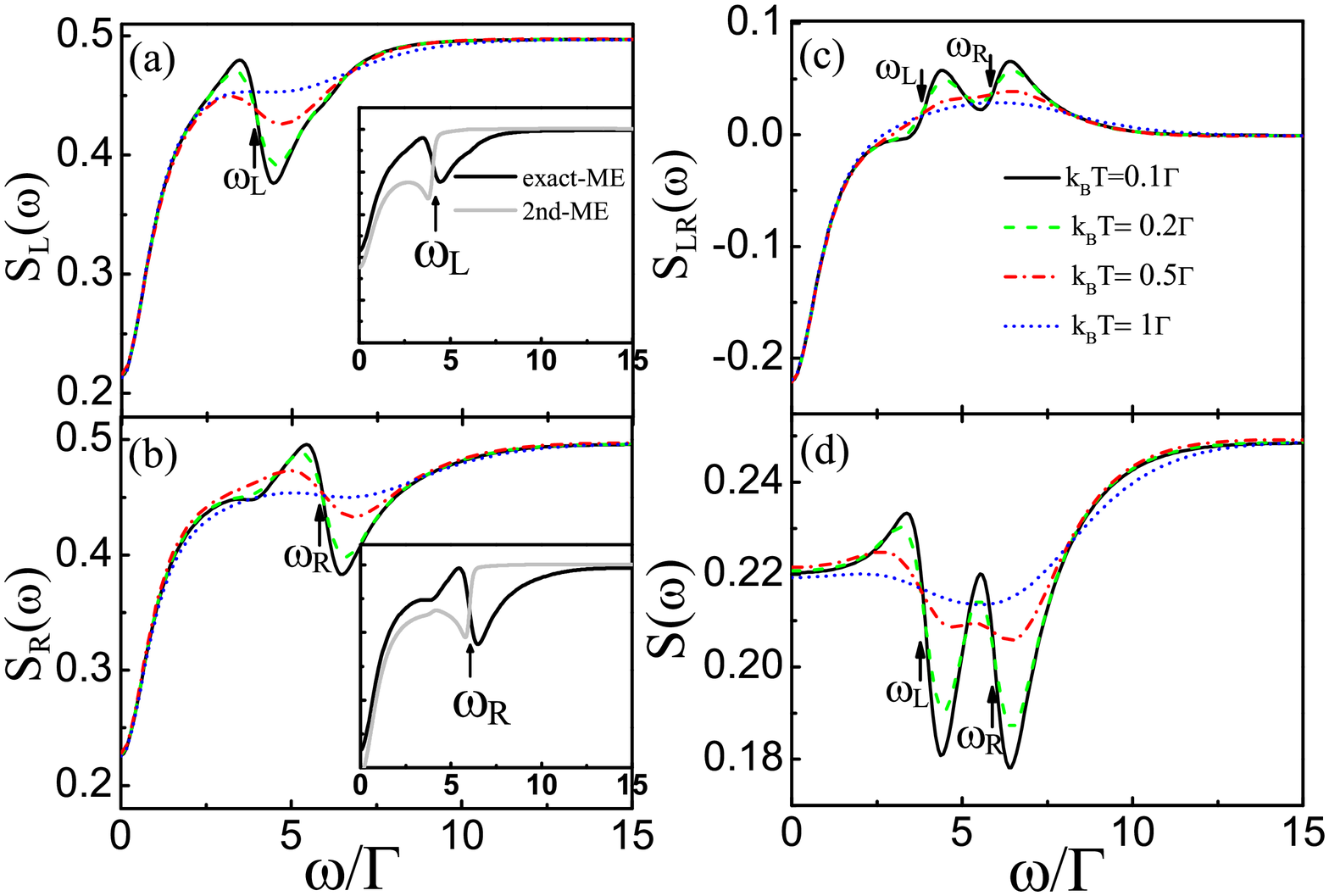}
 \caption{
  Noise spectrum (in unit of $e^2\Gamma/\hbar$)
for current transport through single quantum dot, where
$\epsilon=\Gamma$, with $\Gamma_L=\Gamma_R=0.5\,\Gamma$, under the
bias of $V=10\,\Gamma$ at the specified temperatures.
The peak-dips crossing the tunneling resonance frequencies,
$\omega_{\alpha}=|\epsilon-\mu_{\alpha}|$ with $\alpha=L$ and $R$,
are indicated by arrows. The inserting figures in (a) and (b) give a
comparison between the exact solution with the $2^{\rm nd}$-order
perturbation result at the temperature $k_B T=0.1 \Gamma$.}
 \label{fig1}
\end{figure}

Figure \ref{fig1} shows the current noise spectrums
for auto-correlations
[\Fig{fig1}(a) and (b)], cross-correlation [\Fig{fig1}(c)],
 and circuit current correlation [\Fig{fig1}(d)],
at different temperatures. In the low or high frequency region, the
noise spectrum is nearly independent of temperature, despite of the
fact that the stationary current increases as temperature decreases.
In the large frequency regime, the noise spectrum at the individual
lead $S_{\alpha}(\omega)\rightarrow \Gamma_{\alpha}$, while that of
cross-lead $S_{LR}(\omega)\rightarrow 0$. More interestingly we
observe that a peak-dip feature is developed in the noise spectrum
of current auto- and cross-correlations, crossing each resonant
frequency $\omega_\alpha=|\varepsilon-\mu_\alpha|$ (indicated by an
arrow in \Fig{fig1}) as the temperature decreasing.

The above observations can be understood as follows. As the
zero-frequency noise and also the stationary current are concerned,
the transport system studied here with $\mu_L>\varepsilon>\mu_R$ is
dominated by sequential tunneling processes. In the high frequency
($\omega \gg |\varepsilon-\mu_\alpha|$) region, we have
$S_{LR}(\omega) \rightarrow 0$, due to the electron correlation
between different leads is yet to be established. However,
$S_\alpha(\omega)\rightarrow \Gamma_\alpha$ coming from the fact
that the fluctuations arises mainly from the reservoir background.
Therefore, we cannot find the system-associated structure in the
noise spectrum at both limits. 
However, the shot noise over the full frequency range describes
various tunnelings associated with energy emissions and absorbtions
at different detection frequency $\omega$.
Therefore it must manifest the energy structure dependence of the
dot system through the applied bias voltage and temperature.
As one can see from \Fig{fig1}, the noise spectrum at relative high
temperature shows a smooth energy-structure dependence near the
resonant frequency $\omega_{\alpha}=|\varepsilon-\mu_\alpha|$. This
result is consistent with that of Refs.\,\onlinecite{Eng04136602}
and \onlinecite{Jin11053704}, in which one used the second-order
time-nonlocal master equation which contains very little memory
 when $k_BT>\Gamma$ and describes only the sequential
tunneling process where the Markovian dynamics is dominated.
However, non-Markovian dynamics should play an important role in the
noise spectrum at low temperature ($k_BT\ll \Gamma$) where
cotunneling processes should become significant, especially around
the resonant frequency where the electron can dramatically tunnel
forth and back between the leads and the dot near the Fermi surface
of the leads. The observed peak-dip crossing the resonant frequency
in $S_{\alpha}(\omega)$ [\Fig{fig1}(a) and (b)] and $S_{LR}(\omega)$
[\Fig{fig1}(c)] is indeed such a non-Markovain result as a
combination of sequential tunneling and cotunneling processes at low
temperature. In particular, the positive peak-dip feature in the
cross-correlation fluctuations spectrum must come from the
cross-lead cotunneling contribution.

The above peak-dips crossing the resonant frequencies in the
noise spectrum have not been shown in the previous studies.\cite{Fli08150601,Luo07085325,Don08033532,%
Eng04136602,Rot09075307,Kie06033312,Agu04206601,Jin11053704} This is
because in Refs.\,\onlinecite{Fli08150601,Luo07085325,Don08033532},
the Markovian treatment has been used, while
Refs.\,\onlinecite{Eng04136602} and \onlinecite{Jin11053704}
considered only the sequential tunneling process at relative high
temperature with the $2^{\rm nd}$-order master equation approach
which is also Markovian dynamics dominated.
Refs.\,\onlinecite{Rot09075307} used B\"{u}ttiker's scattering
matrix approach which is exact at zero frequency but contains
approximation that covers mainly the Markovian dynamics at finite
frequency.\cite{But923807,But91199} To make a comparison, we
calculate the noise spectrum with $2^{\rm nd}$-order perturbation
approximation at low temperature. The result is inserted in
\Fig{fig1}(a) and (b). It shows that the $2^{\rm nd}$-order
perturbation noise spectrum shows a dip only but not a peak-dip. As
it is well-known the $2^{\rm nd}$-order perturbation approximation
is invalid at low temperature.
Very recently, using the nonperturbative hierarchical equation of
motion treatment,\cite{Wan12} this peak-dip feature in the noise
spectrum is also observed. \Fig{fig1} also clearly shows that with
the temperature increasing, the aforementioned non-Markovian effect,
i.e. the peak-dips in the noise spectrum, is diminished. We thus
conclude that the observed peak-dip feature in the noise spectrum is
a non-Markovian memory effect through various tunneling processes,
including both the sequential tunneling and cotunneling events.

\subsection{Double quantum dots}

\begin{figure}
\includegraphics*[width=0.9\columnwidth,angle=0]{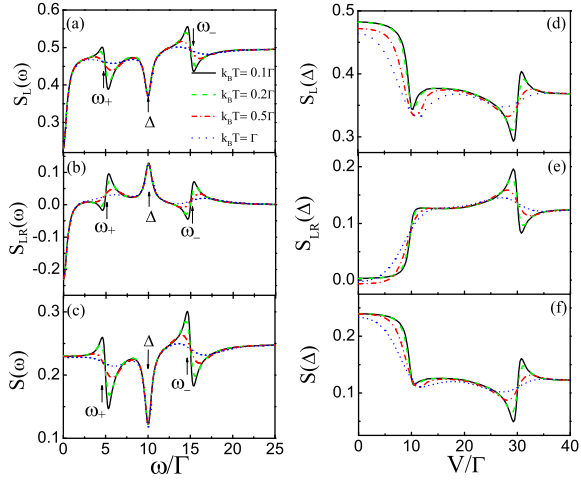}
\caption{ 
 Shot noise (in unit of $e^2\Gamma/\hbar$) of
a double-dots system, where $\varepsilon_l=\varepsilon_r=0$ and
$\Omega=5\,\Gamma$, with $\Gamma_L=\Gamma_R=0.5\,\Gamma$ at the
specified temperatures. Left panels: The noise spectrum as a
function of frequency, under the bias voltage $eV=20\,\Gamma$. The
tunneling resonances of the peak-dip chacteristics occurs at
$\omega_{\pm}=|\varepsilon_\pm-\mu_\alpha|=5\,\Gamma$ and
$15\,\Gamma$ and the Rabi resonance at $\Delta=10\,\Gamma$,
indicated by the arrows individually. Right panels: The shot noise
at the Rabi frequency as function of bias voltage. The
double-resonance $\omega=\Delta=|\varepsilon_{\pm}-\mu_{\alpha}|$
occurs at both $eV=10\,\Gamma$ and $eV=30\,\Gamma$. The former
satisfies also the resonant transport conditions
$\mu_L=\varepsilon_+$ and $\mu_R=\varepsilon_-$, but the latter is
of $\mu_L>\varepsilon_{\pm}>\mu_R$. } \label{fig2}
\end{figure}

Consider now the transient transport through a system of two coupled
quantum dots, described by
$H_S=\sum_{\mu=l,r}\varepsilon_{\mu}d^\dg_{\mu}d_{\mu}
+\Omega\big(d^\dg_{l} d_{r}+d^\dg_{r} d_{l}\big)$. This system has
been studied widely as a charge qubit
\cite{Agu04206601,Luo07085325,Kie07206602,Tu08235311} and has also
been proposed as an alternative detector of a charge
qubit.\cite{Gil06116806}
The intrinsic coherent Rabi frequency $\Delta$ is
the energy difference between eigenstates ($\varepsilon_{\pm}$),
e.g., $\Delta=\varepsilon_{+}-\varepsilon_{-}=2\Omega$ for
the degenerate double-dots system
considered for demonstrations.
It is known\cite{Luo07085325,Agu04206601} that the Rabi coherence of
the central system shows a dip at $\omega=\Delta$ in the
auto-correlation noise spectrum $S_{\alpha}(\omega)$, as can be seen
in \Fig{fig2}(a). But, it appears a peak in the cross-correlation
noise spectrum $S_{LR}(\omega)$, as shown in \Fig{fig2}(b), while
remains a dip in the total circuit noise spectrum, for symmetric
junction capacitances, as depicted in \Fig{fig2}(c).
The above Rabi coherence signatures are nearly independent of the
temperature. Physically, the Rabi coherence is intrinsic. Therefore,
it can be extracted even in the weak system-reservoir coupling
regime, where the $2^{\rm nd}$-order master equation is applicable.
\cite{Luo07085325,Agu04206601}
Besides the coherent signals of Rabi frequency in \Fig{fig2}(a)-(c)
where $\mu_L > \varepsilon_{\pm}> \mu_R$, the expected peak-dips of
non-Markovian characteristics occur at tunneling resonance of
$\omega_\pm=|\varepsilon_{\pm}-\mu_\alpha|$ at low temperature. 
As temperature increases, this peak-dip feature is diminished, just
the same as that in the single quantum dot case described earlier.

Figure \ref{fig2}(d)-(f) depict the noise spectrum at Rabi resonance
$\omega=\Delta$ 
versus the applied bias voltage $V$ that
monitors the
tunneling resonance of $\omega=|\varepsilon_{\pm}-\mu_\alpha|$.
The double-resonance of
$\omega=\Delta=|\varepsilon_{\pm}-\mu_\alpha|$ occurs at
$V=10\,\Gamma$ and
$V=30\,\Gamma$ for the present system in study.
Apparently, the bias voltage dependence of the noise $S(\Delta)$ at
Rabi frequency is changed dramatically at low temperature,
especially at tunneling resonance, as an expected non-Markovian
effect. In particular, at $V=30\,\Gamma$, both the two dot levels
lie within the applied bias window, i.e.,
$\mu_L>\varepsilon_{\pm}>\mu_R$. The resulting peak-dip tunneling
resonance shows up clearly in \Fig{fig2}(d)-(f).
%
The behavior of $S(\omega=\Delta)$ around $V=10\,\Gamma$ is an
interplay between the Rabi resonance and the lead-dot tunneling
resonance.

\begin{figure}
\includegraphics*[width=0.9\columnwidth,angle=0]{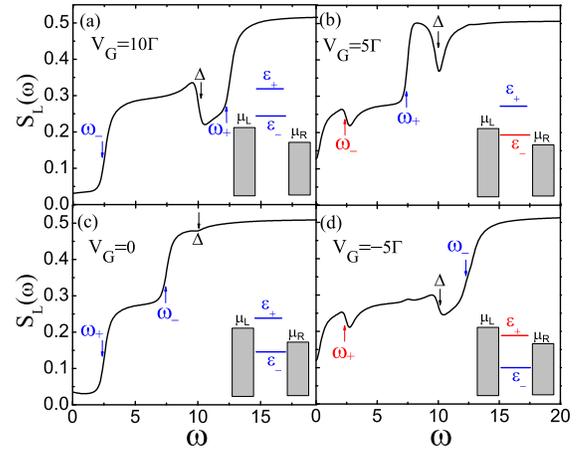}
\caption{Frequency-dependent noise spectrum of the left lead for the double-dots
transport system with different tunneling regimes through gate voltage $V_G$,
i.e., $\varepsilon_l=\varepsilon_r=V_G$, at low temperature $k_BT=0.1\,\Gamma$
with the applied bias voltage
$eV=5\Gamma$. The other parameters are the same as in Fig.\,\ref{fig2}.
}
\label{fig3}
\end{figure}
The frequency-dependent noise spectra demonstrated in \Fig{fig1} and
\Fig{fig2}(a)-(c) are operated in the so-called large bias region of
$\mu_L>\varepsilon,\varepsilon_{\pm}>\mu_R$, where the tunneling
resonance shows the peak-dip feature at low temperature, as a
combination effect of the sequential tunneling and cotunneling
events.
Since the Rabi resonance spectral profile does not depend on the
applied gate voltage for its overall intensity, we shall also
examine the tunneling resonance behavior for different set up of the
applied gate voltage to the dot energy levels. Without loss the
generality, we take a fixed bias voltage, say $V=5\,\Gamma$, and
apply several different gate voltages to the dot levels to achieve
different tunneling scenarios. The results are summarized in
\Fig{fig3}. \Fig{fig3}(a) and (c) describe the case of both the two
dot levels lying outside the bias voltage window. The corresponding
tunneling resonance shows only with a step shape. In contrast, if
one of the dot levels, i.e., the level $\varepsilon_-$ and
$\varepsilon_+$ in \Fig{fig3}(b) and (d), respectively, lies inside
the the bias voltage window, the corresponding tunneling resonance
spectrum generates a peak-dip. As it has been understand, for the
dot level lying outside the bias voltage window (i.e. either
$\varepsilon > \mu_{L,R}$ or $\varepsilon < \mu_{L,R}$), the
sequential tunneling through this level is largely suppressed and
the transport is dominated by cotunneling events. So the cotunneling
transport alone only shows a step near the resonance frequency. The
peak-dip occurs only for $\mu_L>\varepsilon>\mu_R$ at low
temperature, where both the sequential tunneling and cotunneling
participate in the non-Markovian dynamics.

\section{Summary}
\label{thsum} In summary, based on the exact master equation for
nonequilibrium transport, we constructed the corresponding
particle-number resolved master equation, from which we have also
established the formalism for the noise spectrum in the full
frequency range. This new formulism is applicable to arbitrary bias
voltage, temperature and system-reservoir coupling strength for all
the non-Markovian processes in various nanoelectronic systems where
the electron-electron interaction has been ignored.
We applied this formalism to two widely studied transport model
systems, i.e, electron transport through single quantum dot with a
resonant level and double coupled quantum dots containing one energy
level in each dot contacted the electron reservoirs, respectively.
We showed different tunneling characteristics in the noise spectrum
in both the large and small bias voltage regime at different
temperature.
In particular, we found peak-dips in the noise spectrum crossing the
tunneling resonant frequencies which are defined individually by the
energy difference between the applied chemical potentials and the
dots energy levels. This peak-dip feature has not been observed in
previous theoretical works.
The characteristic peak-dip in the current noise spectra is a non-Markovian
effect at low temperature involving both sequential tunneling and
cotunneling. As the temperature increasing (($k_B T>\Gamma$),
peak-dip profile in the noise spectrum is diminished.
In contrast with the aforementioned external bias voltage regulated
resonant characteristics, the internal coherent Rabi oscillation
signal in the double dot system is rather independent of
temperature. The coherent Rabi oscillation results in just a normal
dip profile in the auto-correlation noise spectrum but a peak in the
cross-correlation spectrum.
We expect that these characteristics in the current noise spectra we found
here would be tested in experiments or in other theoretical
calculations.

We should also point out that the characteristic structure showing
peak-dips in the noise spectrum may be changed after the
electron-electron interaction is taken into account.
More interesting phenomena in the noise spectrum should be expected
in the interacting systems at low temperature.
A closed formulation for the exact master equation of the reduced
density matrix and the exact calculation of noise spectrum with the
consideration of Coulomb interaction is not obvious.
However, the present formalism is easy to be extended for including
the Coulomb interaction with respect to the saddle point
approximation \cite{Kam992218} or loop expansion, \cite{Zha921900}
where the Coulomb interaction can be treated self-consistently in
generalizing integrodifferential equations of motion  \Eq{uv-eq}, or
by means of Hierarchical equation of motion approach.
\cite{Jin08234703} 
The work along this line is in progress.


\acknowledgments Support from NNSF China (10904029 and 10905014) and
ZJNSF China (Y6090345), the NSC (NSC-99-2112-M-006-008-MY3) and
National Center for Theoretical Science of Taiwan, and also UGC
(AOE/P-04/08-2) and RGC (604709) of Hong Kong SAR Government is
acknowledged.

\appendix
\section*{Appendix: Numerical method}
\label{thapp-num}
As mentioned in \Sec{thnoise}, the key to the
practical calculation of the noise spectrum as well as the reduced
density matrix \Eq{QME_LB} and current in \Eq{curr} in the present formulism
relies on how to solve the integrodifferential equations of
Eq.~(\ref{uv-eq}), which contain all the non-Markovian memory
effects of the central system interacting with its environment. In
general, it is not possible to analytically solve Eq.~(\ref{uv-eq})
but the numerical solution of such integrodifferential equation is
also very difficult due to the non-Markovian memory kernels.
Here, in terms of a parameterized scheme,
\cite{Jin08234703,Cro09245311} we develop a numerical method in
terms of a closed set of coupled differential equations of motion
for solving $\bm u(t)$ and $\bm v(t)$ to overcome the difficulties
in the direct numerical calculation of the integrodifferential
equations.

For the sake of generality, we start with the energy dependence
spectral density as a Lorentzian-type form centered at
$\epsilon_\alpha$:\cite{Mac06085324,Jin08234703,Tu08235311}
\begin{align}
\Gamma_{\alpha}(\omega)=\frac{\Gamma_{\alpha}W^2_\alpha}
{(\omega-\epsilon_\alpha)^2+W^2_\alpha},
\end{align}
where $\Gamma_{\alpha}$ describes the coupling strength and
$W_\alpha$ is the line width of the source (drain) reservoir with
$\alpha=L (R)$. Obviously the wide band limit (flat band),
$\Gamma_{\alpha}(\omega)=\Gamma_{\alpha}$, is achieved by simply
letting $W_\alpha \rightarrow \infty$. In terms of \Eq{ggbeta}, the
non-local time correlation functions
can be parameterized as \cite{Jin08234703, Jin10083013}
\bsube
\label{corr-t1}
\begin{align}
\bm g_{\alpha}(t,\tau)& =\frac{\Gamma_{\alpha}W_\alpha}{2}
e^{-\gamma_{\alpha 0} (t-\tau)},
\label{gt1}
\\
\widetilde{\bm g}_{\alpha}(t,\tau) &=\sum^M_{m=0} \bm\eta_{\alpha
m}e^{-\gamma_{\alpha m} (t-\tau)}\equiv \sum^M_{m=0} \widetilde{\bm g}_{\alpha m}(t-\tau),
\label{gbetat1}
\end{align}
\esube
with $\widetilde{\bm g}_{\alpha m}(t-\tau)\equiv \bm\eta_{\alpha m}e^{-\gamma_{\alpha m} (t-\tau)}$.
The first term in $\widetilde{\bm g}_{\alpha}(t-\tau)$ with $m=0$ arises from the
pole of the spectral density function, with
\begin{align}
\eta_{\alpha 0}
=\frac{\Gamma_{\alpha}W_\alpha/2}{1+e^{-i\beta_\alpha(\gamma_{\alpha0}-i\mu_\alpha)}}~,
~~ \gamma_{\alpha 0}=W_\alpha+i\epsilon_\alpha.
\end{align}
The other terms with $m>0$ ($M\rightarrow\infty$ in principle)
arise from the Matsubara poles, where the relevant parameters are
explicitly given as
\bsube
\begin{align}
\bm\eta_{\alpha m}&= \frac{i}{\beta_\alpha}\bm\Gamma_{\alpha
}(-i\gamma_{\alpha m}), ~~m=1,\cdots \infty;
\\
\gamma_{\alpha m}&=\frac{(2m-1)\pi}{\beta_\alpha}+i\mu_\alpha.
\end{align}
\esube
The above paramterization is based on the Matsubara frequencies
decomposition of Fermi distribution. An alternative efficient
parameter scheme proposed in Ref.\,\cite{Hu10101106} is
Pad$\acute{e}$ spectrum decomposition for its convergence
significantly faster than other schemes at all temperatures. The
resulting formalisms of \Eq{corr-t1} are unchanged, but the
coefficients for $m>0$ in $\widetilde{\bm g}(t,\tau)$ are modified
accordingly.\cite{Hu10101106} Here, following the procedure in
Refs.\,\onlinecite{Jin08234703,Ale09245311} with the above parameterized
scheme, we will present a numerical method in terms of a closed set
of coupled differential equations of motion, instead of the
integrodifferential equation of \Eq{ut-eq}, to solve the
time-dependent transport current and the noise spectrum.

Introducing a new function
$ \bm {g}^u_\alpha(t)\equiv\int^t_{t_0}d\tau\,\bm g_\alpha(t,\tau)\bm u(\tau)$
which satisfies the following equation
 \be\label{walpha}
 \dot{\bm
{g}}^u_\alpha(t)=\bm g_\alpha(0) \bm u(t) - \gamma_{\alpha0}\bm
{g}^u_\alpha(t), \ee with $\bm
g_\alpha(0)=W_\alpha\bm\Gamma_{\alpha}/2$, then \Eq{ut-eq} is
reduced to
\be\label{ut}
\dot{\bm u}(t)=-i\bm\epsilon(t){\bm
u}(t)-\sum_\alpha\bm {g}^u_\alpha(t).
\ee In other words, the
integrodifferential equation of \Eq{ut-eq} is transformed into a
coupled pure differential equations of \Eq{walpha} and \Eq{ut}, from
which it is rather easy to obtain the time-dependent solution of
$\bm u(t)$ and $\bm {g}^u_\alpha(t)$. On the other hand, the formal
solution of \Eq{vt-eq} is
\begin{align}
 {\sf v}(\tau,t)&=\sum_\alpha\!
   \int^\tau_{t_0}\!\! d\tau_1\!\!\int_{t_0}^{t }\!\!d\tau_2\,
    {\bm u}(\tau, \tau_1)
     \widetilde{\bm g}_\alpha(\tau_1, \tau_2)
    {\bm u}^\dag(t,\tau_2), \label{vt1}
\end{align}
from which we have
\begin{align}\label{vt0}
 \bm v(t)\equiv{\sf v}(t,t)&=\sum_\alpha\!
   \int^t_{t_0}\!\! d\tau_1\!\!\int_{t_0}^{t }\!\!d\tau_2\,
    {\bm u}(t, \tau_1)
     \widetilde{\bm g}_\alpha(\tau_1, \tau_2)
    {\bm u}^\dag(t,\tau_2).
\end{align}
Its time derivation is
\begin{align}\label{vt}
\dot {\bm v}(t)&=-i[\bm\epsilon(t), \bm v(t)]
-\sum_{\alpha}\big[{\cal I}_\alpha(t)+{\rm H.c.}\big]
\end{align}
with ${\cal I}_\alpha(t)={\cal K}_{\alpha}(t)+{\cal Q}_{\alpha}(t)$,
where
\bsube
\begin{align}
{\cal K}_{\alpha}(t)&\equiv
\int_{t_0}^{t}\!d\tau\,
   \bm  g_{\alpha}(t,\tau) {\sf v}(\tau,t),
   \label{calKalpha}
   \\
{\cal Q}_{\alpha}(t)&\equiv -\int_{t_0}^{t}\!d\tau\,\widetilde{\bm
g}_{\alpha}(t,\tau)\bm { u}^\dg(\tau,t) = \sum^M_{m=0}{\cal
Q}_{\alpha m}(t). \label{Jalpha2}
\end{align}
\esube
The second identity in \Eq{Jalpha2} is based on the parameterization
scheme of $\widetilde{\bm g}_{\alpha}(t,\tau)$ given by
\Eq{gbetat1}, which leads to \be {\cal Q}_{\alpha m}(t)
=-\int_{t_0}^{t}\!d\tau\,\widetilde{\bm g}_{\alpha m}(t,\tau) \bm {
u}^\dg(\tau,t), \ee
Furthermore, by introducing new functions
\bsube
\begin{align}
 {\cal C}_{\alpha\alpha'm}(t)&\equiv
\int_{t_0}^{t}\!d\tau\,\int^\tau_{t_0} d\tau_1
  {\bm  g}_{\alpha}(t,\tau){\sf u}(\tau, \tau_1)\widetilde{\bm  g}_{\alpha'm}(\tau_1,t),
 \\
 \label{calDalpha}
{\cal D}_{\alpha\alpha'}(t)&\equiv
\int_{t_0}^{t}\!d\tau\,\int^t_{t_0} d\tau_1
   \bm  g_{\alpha}(t,\tau){\sf v}(\tau, \tau_1)\bm
   g_{\alpha'}(\tau_1,t),
\end{align}
\esube
we can find the following coupled differential equations:
\bsube \label{dd}
\begin{align}
 \dot  {\cal K}_{\alpha}(t)&=
{\cal K}_{\alpha}(t)[i\bm\epsilon(t)-\gamma_{\alpha0} ]
    +  \bm g_\alpha(0)\bm v(t)
    \nl&\quad
  +\sum_{\alpha'm}
\big[{\cal C}_{\alpha\alpha'm}(t)-{\cal D}_{\alpha\alpha'}(t)\big],
\label{calK}
\\
 \dot{ \cal Q}_{\alpha m}(t)&={ \cal Q}_{\alpha m}(t)[i\bm\epsilon(t)
 -\gamma_{\alpha m}] -\bm \eta_{\alpha m}
 +\sum_{\alpha'}{ \cal C}^\dg_{ \alpha'\alpha m}(t),
 \label{calQalpham}
\\
\dot{\cal C}_{\alpha\alpha'm}(t)&=
   -\bm  g_{\alpha}(0){\cal Q}^\dg_{\alpha'm}(t)
   -[\gamma_{\alpha0}(t)+\gamma^\ast_{\alpha' m}(t)]{\cal C}_{\alpha\alpha'm}(t),
   \label{calCalpham}
   \\
\dot{\cal D}_{\alpha\alpha'}(t)&=
\bm  g_{\alpha}(0){\cal K}^\dg_{\alpha'}(t)
+{\cal K}_{\alpha}(t)\bm g_{\alpha'}(0)
\nl&\quad
-[\gamma_{\alpha0}(t)+\gamma^\ast_{\alpha'0}(t)]{\cal D}_{\alpha\alpha'}(t).
\label{calD}
\end{align}
\esube Solving numerically the coupled pure differential equation of
\Eq{dd} together with \Eq{vt}, we thus have the solution of $\bm
v(t)$ and ${\cal I}_\alpha(t)$  without directly calculating the
multi-integrals in \Eq{vt1}. Consequently, all the time-dependent
coefficients in the transient current of \Eq{curr}, the master
equation of \Eq{QME_LB} and the $n$-resolved master equation of
\Eq{nME}, as well as the frequency-dependent noise spectrum of
\Eq{Sw-exact} can be easily obtained through the relations
$\bm\kappa_\alpha(t)=\bm {g}^u_\alpha(t)[\bm u(t)]^{-1}$ and
$\bm\lambda_\alpha(t)={\cal I}_\alpha(t)-\bm\kappa_\alpha(t)\bm
v(t)$.  This numerical method largely simplifies the numerical
difficulties in solving the integrodifferential equations of
\Eq{uv-eq} which involve very complicated non-Markovian memory
kernels.

In the wideband limit $W_\alpha\rightarrow\infty$, the numerical
calculation can be largely simplified. Explicitly, the
temperature-independent memory kernel of \Eq{gt1} is reduced to a
delta function given by $\bm
g(t,\tau)=\frac{\bm\Gamma_\alpha}{2}\delta(t-\tau)$. For the
temperature-dependent memory kernel in \Eq{gbetat1}, the first term
($m=0$) is also reduced to a delta function but the other terms
($m\geq1$) are apparently changed not too much:
\begin{align}
\label{gambeta2}
 \widetilde{\bm g}_{\alpha}(t-\tau) &\rightarrow
\frac{\bm\Gamma_{\alpha}}{2(1+e^{-i\beta_\alpha W_\alpha})}
\delta(t-\tau)
\nl &\quad
 +\frac{i}{\beta_\alpha} \bm\Gamma_{\alpha}\sum^M_{m=1}
e^{-\gamma_{\alpha m} (t-\tau)}.
\end{align}
Thus the non-Markovian effect in the wideband limit is determined by
the temperature together with the bias voltage and the
system-reservoir coupling. If we take further a high temperature
limit $\beta_\alpha\rightarrow 0$, the summation term in
\Eq{gambeta2} will also be reduced to a delta function of $t-\tau$.
Then no any memory effect remains, and a true Markov limit is
reached at high temperature limit.
In the practical numerical calculation, we keep the
temperature-dependent memory kernel with the expression of
\Eq{gbetat1} and the WBL is taken by setting $W_\alpha\ge100\Gamma$.
\cite{Wan07155336,Jin10083013} The equations of the $m$-related
quantities, such as $\bm v(t)$, ${\cal Q}_{\alpha m}(t)$, and ${\cal
C}_{\alpha\alpha'm}(t)$ given by \Eq{vt}, \Eq{calQalpham} and
\Eq{calCalpham}, respectively, are thus not changed.
The other equations can be simplified as follows. The equation of
\Eq{ut} is recast to \be\label{ut2} \dot{\bm
u}(t)=-i\bm\epsilon(t){\bm u}(t)-\frac{\bm \Gamma}{2}\bm u(t), \ee
and \Eq{walpha} is not needed. Also, we can simply solve
\Eq{calKalpha} and \Eq{calDalpha} with the results ${\cal
K}_{\alpha}={\bm\Gamma_\alpha}\bm v(t)/2$ and ${\cal
D}_{\alpha\alpha'}(t)={\bm\Gamma_\alpha}\bm
v(t){\bm\Gamma_{\alpha'}}/4$ but \Eq{calK} and \Eq{calD} are also no
longer used.


\end{document}